\documentclass[aps,prapplied,showpacs,preprintnumbers,reprint,floatfix,superscriptaddress,amsmath,amssymb]{revtex4-2}

\usepackage{graphicx}
\usepackage[linkcolor=blue,citecolor=blue,filecolor=blue,colorlinks=true,urlcolor=blue]{hyperref}

\usepackage{bm}

\usepackage{upgreek}
\usepackage{color}
\usepackage{upgreek}
\usepackage{natbib}

\usepackage{amsmath,amsthm,amsfonts,amssymb,amscd}
\usepackage{lastpage}
\usepackage{enumerate}
\usepackage{fancyhdr}
\usepackage{mathrsfs}
\usepackage{xcolor}
\usepackage{graphicx}
\usepackage{listings}
\usepackage{hyperref}
\usepackage{braket}
\usepackage{amsmath}
\usepackage{mathtools}
\usepackage{lipsum, babel}
\usepackage{float}
\usepackage{array, booktabs,longtable}

\DeclareMathAlphabet{\pazocal}{OMS}{zplm}{m}{n}

\begin{document}

\title{Density-Matrix Model for Photon-Driven Transport in Quantum Cascade Lasers}

\author{S. Soleimanikahnoj}\email{soleimanikah@wisc.edu}
\affiliation{Department of Electrical and Computer Engineering, University
of Wisconsin-Madison, 1415 Engineering Dr., Madison, Wisconsin 53706, USA}

\author{M. L. King}\email{michelle.king@wisc.edu}
\affiliation{Department of Electrical and Computer Engineering, University
of Wisconsin-Madison, 1415 Engineering Dr., Madison, Wisconsin 53706, USA}

\author{I. Knezevic}\email{knezevic@engr.wisc.edu}
\affiliation{Department of Electrical and Computer Engineering, University
of Wisconsin-Madison, 1415 Engineering Dr., Madison, Wisconsin 53706, USA}

\date{\today}

\begin{abstract}
We developed a time-dependent density-matrix model to study photon-assisted (PA) electron transport in quantum cascade lasers. The Markovian equation of motion for the density matrix in the presence of an optical field is solved for an arbitrary field amplitude. Level-broadening terms emerge from microscopic Hamiltonians and supplant the need for empirical parameters that are often employed in related approaches. We show that, in quantum cascade lasers with diagonal design, photon resonances have a pronounced impact on electron dynamics around and above the lasing threshold, an effect that stems from the large spatial separation between the upper and lower lasing states. With the inclusion of PA tunneling, the calculated current density and output power are in good agreement with experiment.
\end{abstract}

\maketitle
\section{Introduction}
Quantum cascade lasers (QCLs) are unipolar sources of coherent radiation emitting in the terahertz and infrared portions of the electromagnetic spectrum~\cite{faist1997laser,kazarinov1971possibility}. The gain medium of a QCL is a periodic stack of compound-semiconductor heterostructures. The resulting multi-quantum-well electron band structure in the growth direction has discrete energy levels, and the associated wave functions are quasibound. While lasing stems from radiative electron transitions between specific states, nonradiative processes mediated by various mechanisms also play important roles in device operation. In particular, photon-assisted (PA) transport becomes significant in terahertz and midinfrared QCLs at and above the lasing threshold~\cite{matyas2011photon,lindskog2014comparative,choi2008gain}.  This effect of the optical field also occurs in superconducting junctions~\cite{tien1963multiphoton}, optical lattices~\cite{sias2008observation}, superlattices~\cite{keay1995photon,wacker2002semiconductor,feldmann1992optical,ignatov1976nonlinear,ktitorov1972bragg}, and other quantum-well structures~\cite{asada2001density}.

In QCLs, PA transport has been modeled using semiclassical methods (rate equations and Monte Carlo ~\cite{choi2008gain,blaser2001characterization,matyas2011photon,faist1997laser}) and quantum-mechanical methods (density matrix and nonequilibrium Green's function  ~\cite{willenberg2003intersubband,kumar2009coherence,weber2009density,terazzi2010density,wacker2002gain,bismuto2010electrically,lindskog2014comparative,lee2002nonequilibrium,bugajski2014mid,wacker2013nonequilibrium,lindskog2014comparative,kolek2014impact,kolek2019implementation}). The modeling efforts reliant on the rate equations for level populations employed empirical or phenomenological scattering rates to characterize PA transport ~\cite{choi2008gain,blaser2001characterization,matyas2011photon}. More recent density-matrix (DM) models achieved quantitative agreement with experimental studies~\cite{lindskog2014comparative,bismuto2010electrically}, however, in these models, the PA scattering rates were calculated using Fermi's golden rule in which the energy-conserving delta functions were replaced by empirically modified Lorentzian terms. The nonequilibrium Green’s function (NEGF) technique allows for a methodical treatment of scattering, but the theoretical complexity and computational demands of NEGF make it inconvenient for use by experimental researchers interested in QCL design and optimization, who rely on the concepts of quasibound states and their lifetimes as the cornerstone of intuition building. Therefore, there is a need for a computationally efficient quantum-transport treatment of PA tunneling in QCLs that does not require phenomenological parameters and that employs broadly adopted intuitive concepts.

 \begin{figure}
  \includegraphics[width=0.7\linewidth]{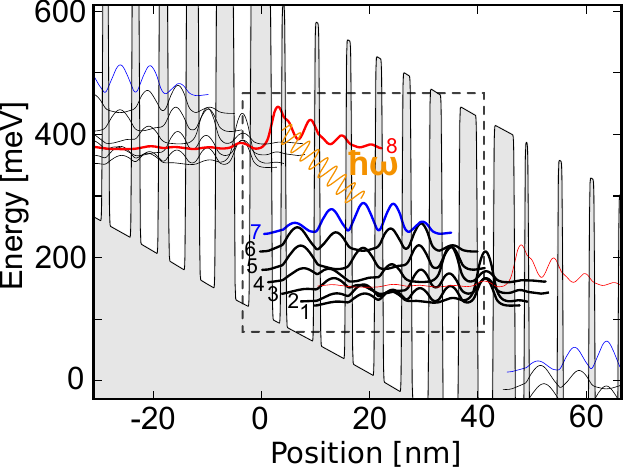}
  \caption{Conduction-band edge and probability densities for the eight eigenstates used in calculations (bold curves) at an above-threshold electric field bias of 50 kV/cm. The states that belong to neighboring periods are denoted by thin gray curves and the dashed box indicates a single stage, starting with the injection barrier. The states are numbered in the order of increasing energy, starting with the ground state; the radiative transition occurs from 8 to 7. The layer structure (in nanometers), starting with the injector barrier (centered at the origin) is \textbf{4.0}/1.8/\textbf{0.8}/5.3/\textbf{1.0}/4.8/\textbf{1.1}/4.3/\textbf{1.4}/3.6/\textbf{1.7}/3.3/\textbf{2.4}/\underline{3.1}/ \underline{\textbf{3.4}}/2.9, with the barriers denoted in bold. Underlined layers are doped to $1.2\times10^{17}$~cm$^{-3}$, which results in an average charge density of $n_\mathrm{3D}=1.74\times10^{16}$~cm$^{-3}$ per stage. }
  \label{Fig1}
\end{figure}

In this paper, we present a quantum-mechanical model for photon-driven transport in QCLs that is computationally inexpensive, requires no phenomenological parameters, and is conducive to intuition building. The model stems from a rigorous theoretical framework with a positivity-preserving Markovian master equation of motion for the density matrix \cite{Oli_Dissertation,jonasson2017density,Jonasson2016_JCEL,Jonasson2016_Photonics}. The equation of motion is solved self-consistently and nonperturbatively. The solution is used to compute the steady-state and frequency response of the midinfrared QCL from Ref. \cite{bismuto2010electrically} (Fig. \ref{Fig1}). Our results show that the inclusion of PA tunneling leads to substantial changes in electron transport around and above the lasing threshold, and explain why those changes are quite pronounced in QCLs with a diagonal design. Specifically, a significant increase in the current density is observed upon inclusion of PA tunneling, which leads to a better above-threshold agreement between the computed and experimental current density versus field curves than obtained from the models that neglect PA phenomena. In addition, the model enables calculation of the optical gain and output power versus current density. The calculated output power is in close agreement with experiment.

\section{Methodology}

The simulated structure is a midinfrared QCL from Ref.~\cite{bismuto2010electrically}, which emits at around $8.5\,\mathrm{\mu m}$. In this structure, electron dynamics is described by a three-part Hamiltonian
\begin{equation}
\label{Eq1}
    H = H_0 + H_i + H'(t).
\end{equation}
Here, $H_0$ is a three-band $\mathbf{k}\cdot\mathbf{p}$ Hamiltonian capturing the electronic band structure in the growth direction ($z$) and free motion in the $x$-$y$ plane. This Hamiltonian has been extensively used in numerical studies of midinfrared QCLs~\cite{Jonasson2016_JCEL,shi2017modeling,shi2014nonequilibrium}. The effect of a \emph{dc}-bias electric field $\pazocal{E}_{dc}$ is incorporated in $H_{0}$. $H_i$ represents the nonradiative interactions of electrons with longitudinal acoustic and optical phonons, interface roughness, ionized impurities, and random alloy disorder~\cite{jonasson2017density,Oli_Dissertation}. $H'(t)$ stands for the interaction between the optical field and electrons and is defined as
\begin{equation}
\label{Eq2}
    H'(t) = \mathrm{q}\pazocal{E}_{ac}z \cos(\omega t),
\end{equation}
where $\mathrm{q}$ is electron charge, $\pazocal{E}_{ac}$ is the electric-field amplitude, $z$ is position in the growth direction, and $\omega$ is the frequency of radiation. Figure~\ref{Fig1} shows the energy levels and probability densities of the eigenstates associated with $H_0$. For the central period, denoted by a dashed rectangle, the eight states with non-negligible contributions to electron transport are shown in bold. The states associated with adjacent periods are represented by thin gray curves. The lasing transition is between the upper (8) and lower lower (7) lasing levels, shown in red and blue, respectively. In this QCL, the lasing transition is diagonal: the upper and lower states are spatially well separated.

Transport quantities of interest are calculated using the density matrix $\rho$. For calculation purposes, we assume that the device area in the $x$-$y$ plane perpendicular to the transport direction is macroscopic and that the in-plane electron dynamics is that of a free particle in a uniform and isotropic two-dimensional medium. The basis used for transport calculations is $\ket{i,k} = \ket{i}\otimes \ket{k}$. $\ket{i}$ labels the eigenstates of $H_0$ and the continuous parameter $\ket{k}$ represents the amplitude of the wave vector $\mathbf{k} = (k_x,k_y)$ in the $x$-$y$ plane. Owing to translational invariance and isotropy for in-plane dynamics, the density matrix is diagonal in $\mathbf{k}$. The density matrix elements $\braket{k,i|\rho|j,k}$ are shown as $\rho_{i,j}^{E_k}$, with $E_{k} = \hbar^2k^2/2m$ being the in-plane energy; the constant in-plane inverse effective mass used here is a weighted average of the layer-specific inverse effective masses \cite{Oli_Dissertation,jonasson2017density}. Here, we present the key derivation steps of the equation of motion for $\rho_{i,j}^{E_k}$; details are given in the appendix. The density matrix is governed by the equation of motion
\begin{equation}
\begin{gathered}
\label{Eq3}
i\hbar\dot{\rho}^{E_{k}}_{i,j}  = \Delta E_{i,j} \rho^{E_{k}}_{i,j} + \sum_{l,p}\left(H'_{i,l}\rho^{E_k}_{l,j} - H'_{p,j}\rho^{E_k}_{i,p}  \right) + i\hbar\left[\pazocal{D}\rho\right]^{E_{k}}_{i,j}.
\end{gathered}
\end{equation}
$\Delta E_{i,j}$ is the energy spacing between states $i$ and $j$. For a given operator $\hat{O}$, $O_{i,j}$ is the short-hand notation for $\braket{i|\hat{O}|j}$.
$\pazocal{D}\rho$, which will be referred to as the dissipator, is a term that captures the effect of nonradiative interactions in the Markovian limit. The map $\pazocal{D}$ should be thought of as the quantum-mechanical generalization of the scattering rate, supplanting the need for phenomenological dephasing times. The derivation of $\pazocal{D}$ for each specific scattering mechanism is explained in detail in Refs. ~\cite{Oli_Dissertation,jonasson2017density}. The matrix element of $\pazocal{D}\rho$ with indices $i$ and $j$ can be written as a sum of the scattering terms proportional to the density matrix term with the same indices, -$\gamma_{i,j}^{E_k}\rho_{i,j}^{E_k}$ ($\gamma_{i,j}^{E_k}>0$ have the meaning of rates), and the reduced dissipator $\left[\bar{\pazocal{D}} \rho\right]_{i,j}^{E_k}$ that comprises terms proportional to other matrix elements of $\rho$, i.e., $\left[\pazocal{D}\rho\right]_{i,j}^{E_k} = -\gamma_{i,j}^{E_k}\rho_{i,j}^{E_k} + \left[\bar{\pazocal{D}}\rho\right]_{i,j}^{E_k}$. This separation makes Eq.~(\ref{Eq3}) a first-order differential equation for $\rho_{i,j}^{E_k}$, in which the other matrix elements of $\rho$ play a role through $\bar{\pazocal{D}}\rho$ term and a polarization term that involves optical-field amplitude $\pazocal{E}_{ac}$ is present:

\begin{equation}
\begin{split}\label{Eq4}
i\hbar\dot{\rho}^{E_\mathbf{k}}_{i,j} & = \left(\Delta E_{i,j}-i\hbar\gamma_{i,j}^{E_\mathbf{k}}
+ q\pazocal{E}_{ac}\left(z_{i,i} - z_{j,j}\right)\cos(\omega t)\right)\rho^{E_\mathbf{k}}_{i,j} \\ &
+ q\pazocal{E}_{ac}\cos(\omega t)\sum_{\substack{l \neq i \\ p \neq j}}\left(z_{i,l}\rho^{E_\mathbf{k}}_{l,j} - z_{p,j}\rho^{E_\mathbf{k}}_{i,p}  \right) + i\hbar\left[\bar{\pazocal{D}}{\rho}\right]^{E_\mathbf{k}}_{i,j}.
\end{split}
\end{equation}

\noindent In order to address high optical-field amplitudes that characterize above-threshold laser operation, we must seek a solution that is nonperturbative in $\pazocal{E}_{ac}$. To that end, we solve Eq. (\ref{Eq4}) iteratively: the last two terms on the right-hand side will use a previous iteration of $\rho$, denoted by $\breve\rho$ here, whereby the equation is turned into a first-order ordinary differential equation with a known general solution (see appendix), which, after some algebraic manipulation, can be written as:

\begin{widetext}
\begin{equation}
\begin{split}\label{Eq5}
& \rho_{i,j}^{E_\mathbf{k}}(t) = \sum_{n,m,q}J_n\left(\frac{V_{i,j}}{h\omega}\right)J_{m}\left(\frac{V_{i,j}}{h\omega}\right)\Biggl\{ -i\hbar\left[\bar{\pazocal{D}}\breve\rho_q\right]^{E_\mathbf{k}}_{i,j}\frac{e^{i (n-m+q)\omega t}}{\Delta E_{i,j} -i\hbar\gamma_{i,j}^{E_\mathbf{k}} + \hbar\omega (n+q)}\\
   &+ \left[\pazocal{A}_q\right]_{i,j}^{E_\mathbf{k}}
\left( \frac{e^{i (n-m+q+1) \omega t}}{\Delta E_{i,j} -i\hbar\gamma_{i,j}^{E_\mathbf{k}} + \hbar\omega (n+q+1)} + \frac{e^{ i(n-m+q-1) \omega t}}{\Delta E_{i,j} -i\hbar\gamma_{i,j}^{E_\mathbf{k}} + \hbar\omega (n+q-1)} \right)\Biggr\}.
\end{split}
\end{equation}
\end{widetext}

Here, $J_{n}$ is the $n$th-order Bessel function of the first kind. $V_{i,j} = q\pazocal{E}_{ac}\left(z_{i,i} - z_{j,j}\right)$. $\breve\rho_{q}$ is the $q$th harmonic component of the previous iteration $\breve\rho(t)$ ($\breve\rho(t) = \sum_{q}\breve\rho_qe^{iq\omega t}$). $\left[\pazocal{A}_q\right]_{i,j}^{E_\mathbf{k}}$,  referred to as the polarization matrix, is defined as

\begin{equation}\label{Eq6}
    \left[\pazocal{A}_q\right]_{i,j}^{E_\mathbf{k}} =  -\frac{\mathrm{q}\pazocal{E}_{ac}}{2}\sum_{\substack{l \neq i \\ p \neq j}}\left(z_{i,l}{[\breve\rho_q]}^{E_\mathbf{k}}_{l,j} - z_{p,j}{[\breve\rho_q]}^{E_\mathbf{k}}_{i,p}  \right).
\end{equation}

\noindent Equations (\ref{Eq4}) and (\ref{Eq5}) describe a driven damped system whose ``steady state'' (the long-time limit of the solution) will be time-dependent. In Eq. (\ref{Eq5}), broadening of the photon resonances is dictated by the nonradiative scattering rates ($\gamma_{i,j}^{E_k}$) calculated directly from microscopic interaction Hamiltonians. This is an improvement over previously developed density-matrix models, where the Lorentzian terms in Fermi's golden rule were assigned an empirical value for the broadening~\cite{bismuto2010electrically,lindskog2014comparative}. In diagonal QCLs, a sharp transition (small $\hbar\gamma_{8,7}^{E_k}$) between the upper and lower lasing levels is expected, because the nonradiative transition rates are low owing to the small spatial overlap between these two states~\cite{kumar2009coherence}. The modulating Bessel term in  Eq.~\ref{Eq5} [$J_n{\left(V_{i,j}/\hbar\omega\right)}J_m{\left(V_{i,j}/\hbar\omega\right)}$] often appears in the calculation of PA transport in superconducting junctions~\cite{tien1963multiphoton}, optical lattices~\cite{sias2008observation}, superlattices~\cite{keay1995photon}, and other quantum-well structures~\cite{asada2001density}. This term determines the severity of PA effects in electron transport. Since $V_{i,j}\sim (z_{i,i} - z_{j,j})$, a larger spatial separation between the upper and lower lasing levels leads to stronger photon-assisted resonances. This is the case in QCLs with diagonal design, where the small spatial overlap between the lasing levels means less parasitic transport between the two lasing levels, but it also leads to strong PA tunneling ~\cite{matyas2011photon,kumar2009186,bai2010quantum,wacker2002gain,wacker2013nonequilibrium}.

The \emph{dc} Fourier component of the solution to Eq. (\ref{Eq5}), denoted $\rho_0$, gives us information on the measured \emph{dc} current density under steady-state operation. $\rho_0$ comprises the terms in which the exponents are zero. Frequency-dependent characteristics of the system can be found from the first harmonic components ($\rho_{\pm1}$) of the density matrix in Eq.~(\ref{Eq5}). $\rho_{+1}$ and $\rho_{-1}$ comprise the terms in which the exponents are $+1$ and $-1$, respectively. Closed-form solutions of $\rho_0$ and $\rho_{\pm1}$ obtained from Eq.~(\ref{Eq5}) are interdependent, and also generally dependent on the Fourier components of higher order ($\{\rho_q : |q| \ge 2\}$). We solve Eq.~(\ref{Eq5}) for the Fourier components of $\rho$ iteratively, considering harmonic components up to and including the second order because we found higher-order components to be numerically insignificant. Convergence is achieved when the $\rho$s on the two sides of Eq.~(\ref{Eq5}) become numerically indistinguishable. A few tens of iterations are typically needed if one starts from a zeroth-order guess that is a simple diagonal density matrix with level populations given by a Fermi distribution on the diagonal (see details in the Appendix). At higher optical fields, more Fourier components may have to be included in the calculation in order to achieve convergence. Furthermore, in the case of terahertz QCLs (which have lower lasing energies and wider wells than their midinfrared counterparts, so the optical-field potential $V_{i,j}$ may exceed the low lasing energy even at relatively low optical fields), the resonances stemming from higher-order harmonic components can have a significant effects on the electronic transport and gain spectrum,~\cite{wacker2013nonequilibrium}, should be included in the calculation.

\section{Results}

\begin{figure}
   \includegraphics[width=0.65\linewidth]{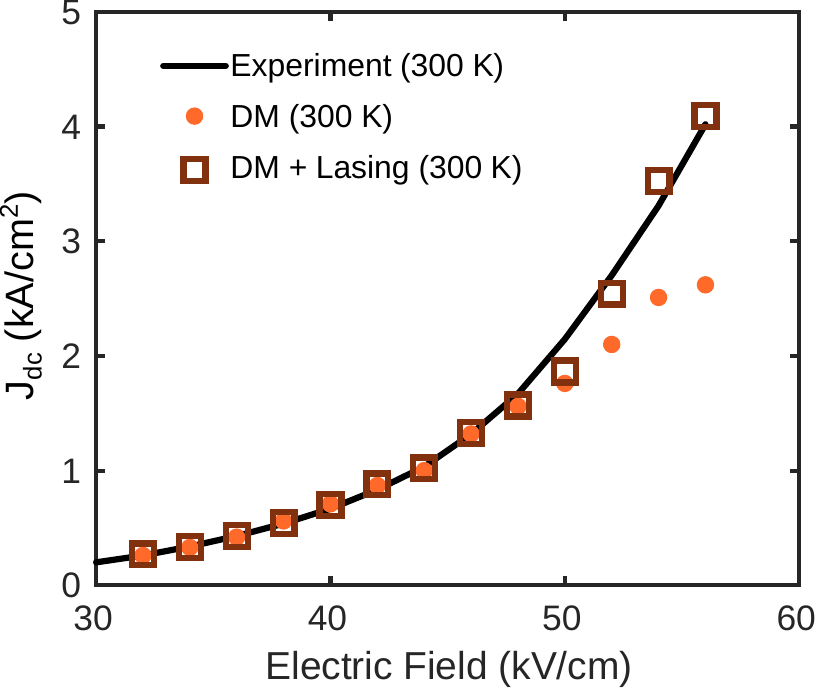}\caption{Calculated \emph{dc} current density as a function of the applied bias electric field at 300 K, with (open squares) and without (solid circles) the influence of photon-assisted tunneling. Experimental data (solid line) is from Fig. 2 in Ref.~\cite{bismuto2010electrically}; the bias electric field was calculated by dividing measured voltage by the total length of the active region (number of periods (50) times period length (44.9 nm)).}
  \label{Fig2}
\end{figure}

Once the density matrix is calculated, the \emph{dc} current density can be written as~\cite{jonasson2017density}
\begin{equation}
\begin{gathered}
\label{Eq7}
J_{dc} = \mathrm{q}n_{3D}\sum_{i,j}\int dE_{k} \;v_{i,j}(\rho_0)^{E_k}_{i,j},
\end{gathered}
\end{equation}
\noindent where $v_{i,j} = \frac{i}{\hbar} \Delta E_{i,j}z_{i,j}$ is the drift velocity matrix. $n_{3D}$, is the average three-dimensional (3D) electron density in the device set to $1.74\times 10^{16}\,\mathrm{m}^{-3}$.

The calculated current density $J_{dc}$ vs. bias electric field $\pazocal{E}_{dc}$ is plotted in Fig.~\ref{Fig2}. $\pazocal{E}_{ac}$ is set to zero for the simulation without lasing. When lasing is included, $\pazocal{E}_{ac}$ is nonzero and its value is set to the operational field at a given bias, as obtained in experiment \cite{bismuto2010electrically}. (The operational field is the optical field at which optical gain reaches lasing threshold, and we use the same values as in Fig.~\ref{Fig5} below.) As seen in Fig.~\ref{Fig2}, the current density increases dramatically when lasing is included  with respect to the simulation without lasing. The increase is more noticeable at higher bias fields, where light amplification takes place. As explained below, this increase can be attributed largely to PA electron transport between the upper and lower lasing levels. The current density with lasing included shows a close agreement with experiment, which underscores the important role of PA tunneling in electron transport. The increase in current following the inclusion of the optical field was observed in previous semiclassical~\cite{matyas2011photon} and quantum-mechanical~\cite{lindskog2014comparative} studies of QCLs and was attributed to PA transport. The same phenomenon was observed in other tunneling structures, as well~\cite{matyas2011photon,lindskog2014comparative,choi2008gain}.

\begin{figure}
\includegraphics[width=1.0\linewidth]{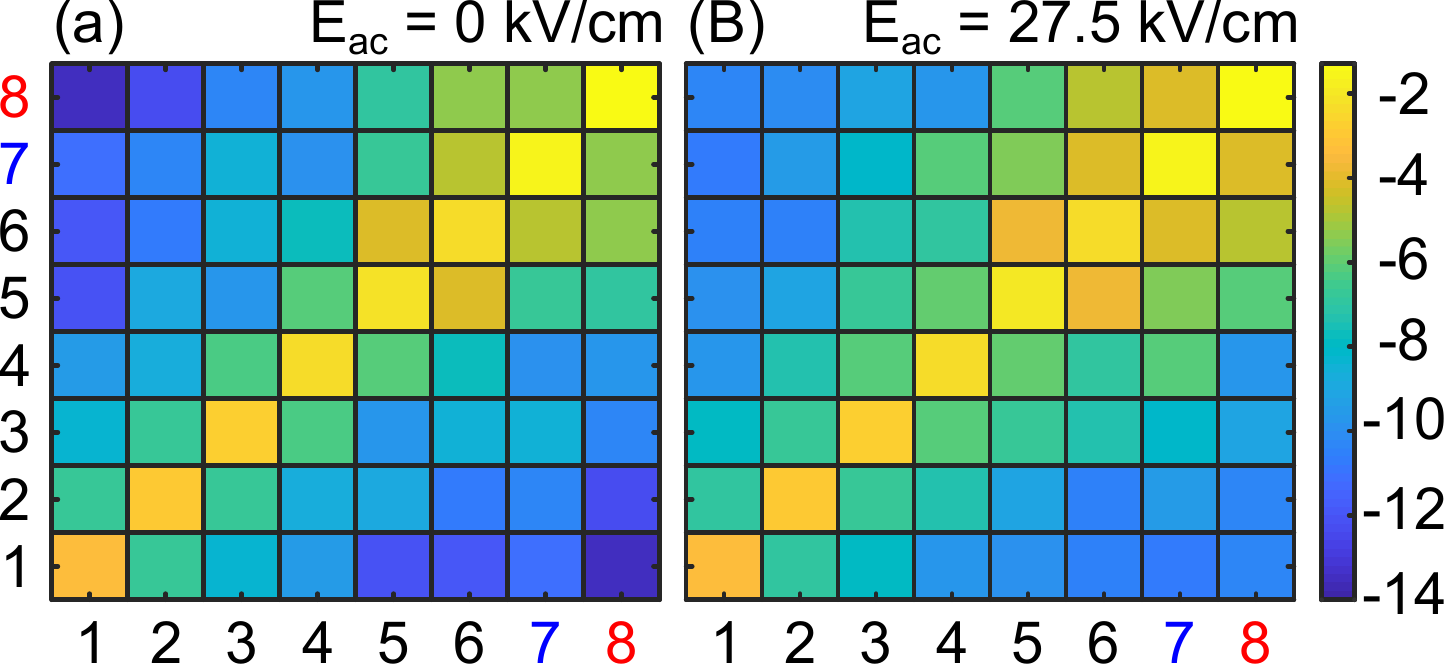}
\caption{The absolute value of static (\emph{dc}) density matrices at $\pazocal{E}_{ac} = 0$ kV/cm (a) and $\pazocal{E}_{ac} = 27.5$ kV/cm (b), after integration over the in-plane kinetic energy. The colorbar scale is logarithmic. The bias electric field is 56 kV/cm (above threshold) and the lattice temperature is 300 K. Upon inclusion of photon-assisted tunneling  the off-diagonal elements (so-called coherences, which play a key role in current flow), increase in magnitude. In particular, the coherence between the upper and lower lasing levels (8 and 7) is enhanced in the presence of the optical field.}
  \label{Fig3}
\end{figure}

According to Eq.~(\ref{Eq7}), only off-diagonal elements of $\rho_0$ contribute to the current density as $v_{i,j} = 0$ for  $i=j$. Therefore, the PA increase of current density observed in Fig.~\ref{Fig2} should be reflected in the off-diagonal elements of $\rho_0$. Figure~\ref{Fig3} shows the log-scale absolute value of the static density matrix ($\rho_0$) with ($\pazocal{E}_{ac} = 27.5$  kV/cm) and without ($\pazocal{E}_{ac} = 0$  kV/cm) lasing. In both panels the static components of the density matrices for the QCL in Fig.~\ref{Fig1} are calculated using Eq.~(\ref{Eq5}) and the in-plane energy ($\tilde{\rho} = \int dE_k\rho^{E_k}$) is integrated in order to obtain a square density matrix. As can be seen, some off-diagonal terms (the so-called coherences) are higher when PA effects are included. Of particular note is the PA enhancement of the coherence between the upper and lower lasing levels. Therefore, the PA increase in the current density largely stems from tunneling between the upper and lower lasing levels. Diagonal elements (population of states) also change with the inclusion of the optical field, but this change is small. This is in agreement with previous studies of diagonal QCLs, where population of lasing states was kept approximately constant while lasing was achieved~\cite{blaser2001characterization,bismuto2010electrically}.

The induced current densities $J_{-1}(\omega)$ and $J_{+1}(\omega)$ are defined in the same way as the static current density in Eq.~(\ref{Eq7}), only $\rho_0$ is replaced by $\rho_{-1}$ and $\rho_{+1}$ respectively. From there, we compute gain $g(\omega)$  as~\cite{wacker2013nonequilibrium}
\begin{equation}
\begin{gathered}
\label{Eq11}
g(\omega) = -\frac{1}{c\epsilon_0\sqrt{\epsilon_r}}\frac{J_{-1}(\omega)+J_{+1}(\omega)}{\pazocal{E}_{ac}}\, ,
\end{gathered}
\end{equation}
where $c$ is the speed of light in vacuum, and $\epsilon_r$ is the background relative permittivity set to the weighted average of the relative permittivities of the composing materials~\cite{Oli_Dissertation,jonasson2017density}.

\begin{figure}
  \includegraphics[width=0.7\linewidth]{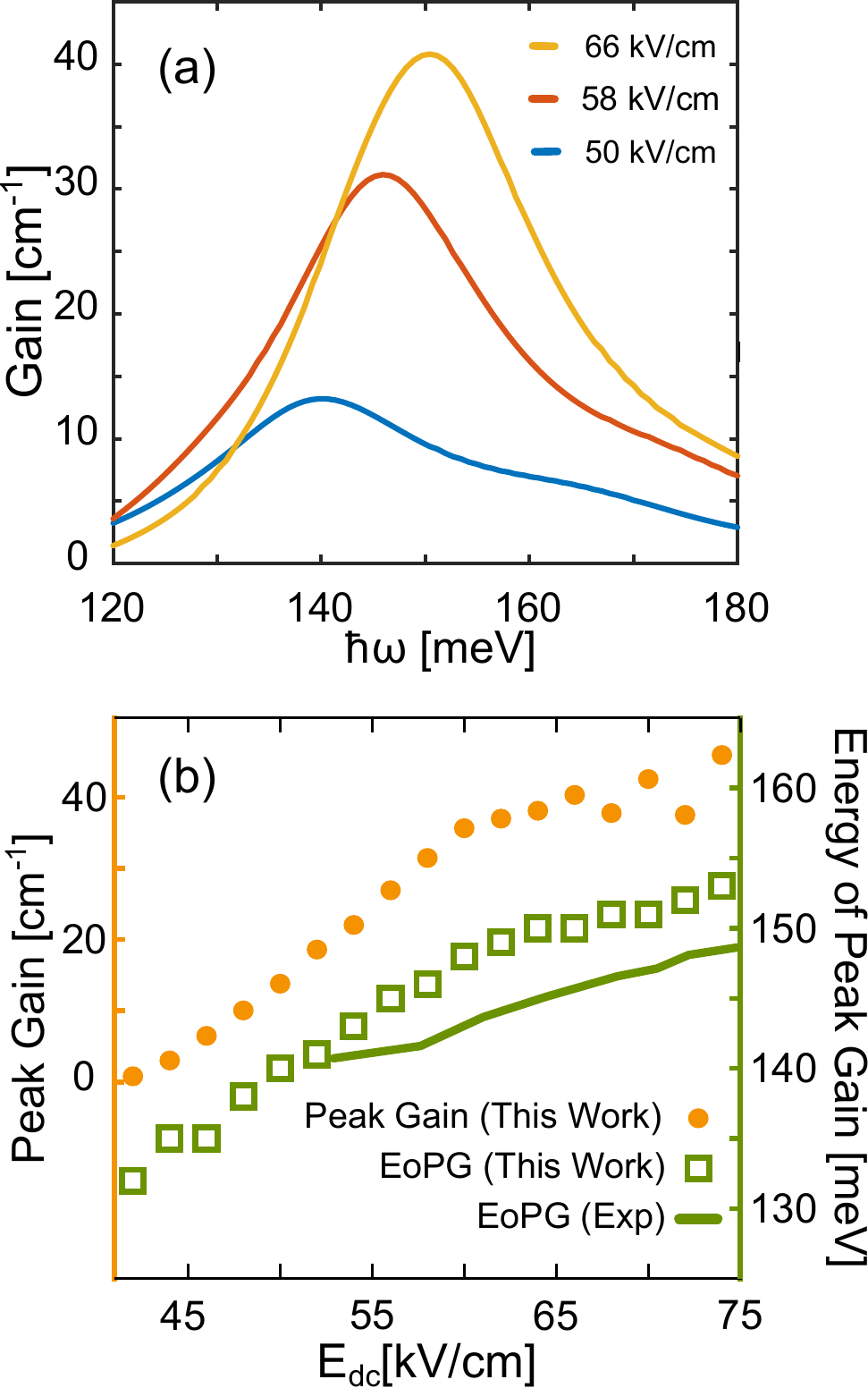}
  \caption{(a) Gain versus photon energy for three different values of the bias electric field $\pazocal{E}_{dc}$. (b) Peak gain (circles) and the photon energy corresponding to peak gain (squares) as a function of the bias electric field. The solid line represents experimental data in Ref.~\cite{bismuto2010electrically} for the photon energy at peak gain. In both panels, gain is calculated in the  limit of vanishing optical field ($\pazocal{E}_{ac} = 0.02$ kV/cm) and the lattice temperature is 300 K.}
  \label{Fig4}
\end{figure}

Figure~\ref{Fig4}(a) shows unsaturated gain as a function of the lasing frequency for three different values of the bias electric field $\pazocal{E}_{dc}$. At each bias, the photon energy $\hbar\omega$ corresponding to peak gain is dictated by the energy difference between the upper and lower lasing levels, i.e., $E_{8,7} = \hbar\omega$. The frequency of peak gain shifts toward higher values as we increase $\pazocal{E}_{dc}$ because of the stronger Stark effect at higher bias fields. Figure~\ref{Fig4}(b) shows peak gain and the photon energy corresponding to peak gain versus the bias electric field. The energy of maximum gain is somewhat higher than the experimentally observed photon energy, possibly owing to the Lamb shift ~\cite{Breuer2002} (the slight interaction-induced change to energy levels), which the calculation neglects. In experiment, the threshold gain is estimated to be $G_{th} \approx 10\,
\mathrm{cm^{-1}}$ at $\pazocal{E}_{dc,th} = 48$ kV/cm~\cite{bismuto2010electrically}. At this bias ($\pazocal{E}_{dc}= 48$ kV/cm), our results show the same value for peak gain (about $10\, \mathrm{cm^{-1}}$).

\begin{figure}
   \includegraphics[width=1.0\linewidth]{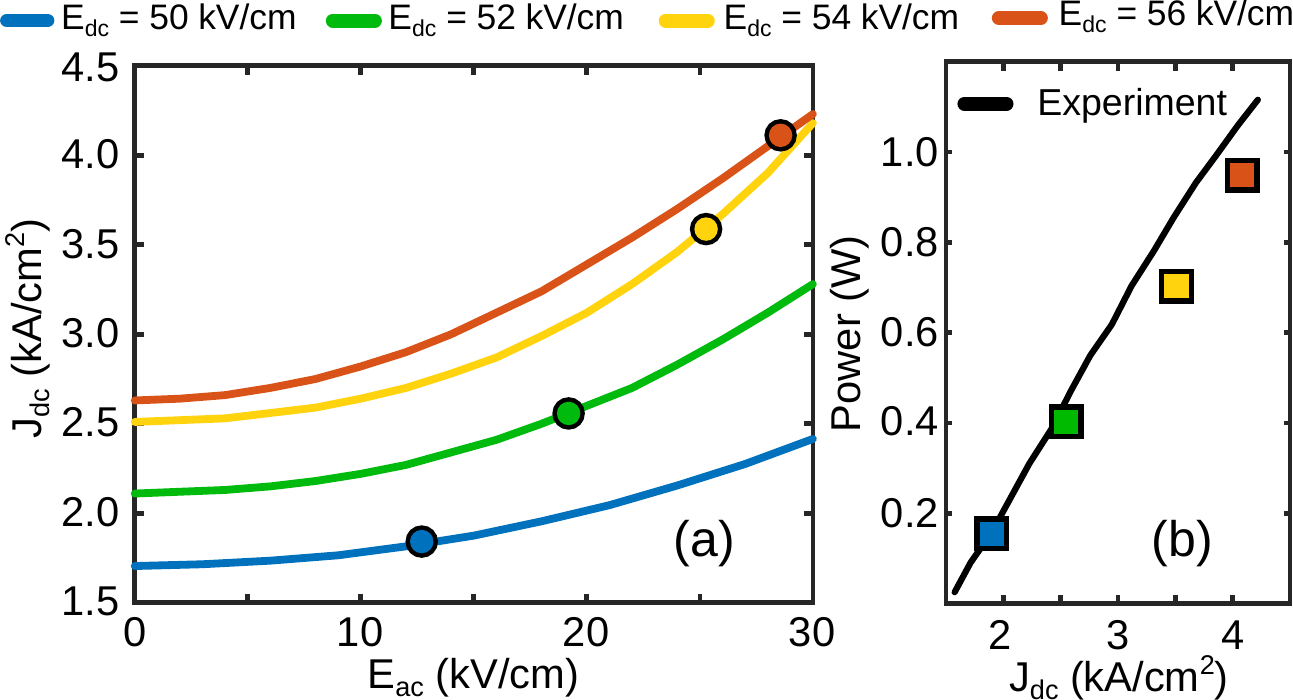}
  \caption{ (a) Calculated dc current density versus optical field at the four dc bias fields given in the legend (top). Circles indicate operating points. (b) Measured (solid line) and calculated (squares) output power as a function of the dc current density. Curves in panel (a) and data points in both panels are color coded to indicate the corresponding bias field in the legend (top). Experimental data is from Ref. \cite{bismuto2010electrically}.}
  \label{Fig5}
\end{figure}

Figure~\ref{Fig5}(a) shows the current density as a function of the optical field for four above-threshold dc bias fields. Operation values of the optical field \cite{bismuto2010electrically} are denoted by circles. (The calculated current density under operation was plotted in Fig.~\ref{Fig2} above, and it is in much better agreement with experiment than the simulations without the laser field.) Figure~\ref{Fig5}(b) shows the resulting output optical power as a function of the current density. Data points are color-coded in order to emphasize the corresponding bias field. Output optical power versus $\pazocal{E}_{ac}$ for bias fields above threshold was calculated by assuming the relation for a traveling wave ($\mathrm{Power} = \pazocal{C}\pazocal{E}_{ac}^2$), where $\pazocal{C}$ is a proportionality constant that depends on design parameters of the QCL waveguide. For the device studied here, an output power of $\approx$ 1 W was reported for $\pazocal{E}_{ac}  = 30$ kV/cm~\cite{bismuto2010electrically}, which implies $\pazocal{C} \approx 1.11^{-13}$ W.m/V. As can be seen, the calculated power agrees well with the experimental data.

\section{Conclusion}
In summary, we developed a theoretical model based on the density matrix to study the effects of the optical field on electron dynamics in QCLs. The model supplants the need for empirical broadening values that characterized earlier density-matrix work. We solved the Markovian equation of motion for the density matrix in the presence of an optical field; the solution is nonperturbative, i.e., holds for any field amplitude. Based on the computed density matrix, we obtained the steady-state and frequency-dependent characteristics of a QCL. We showed that spatial separation between lasing levels is a critical factor in the intensity of PA tunneling. The PA effect on the current density in the simulated QCLs with a diagonal design is quite pronounced. With PA transport included, the agreement between calculation and experiment is excellent.

\section{Acknowledgement}

The authors gratefully acknowledge Dan Botez and Luke Mawst for useful discussions. This work was supported by the DOE-BES award DE-SC0008712 (theory) and by the AFOSR award FA9550-18-1-0340 (simulation). Preliminary work was partially supported by the Splinter Professorship. Calculations were performed using the compute resources of the UW-Madison Center For High Throughput Computing (CHTC) in the Department of Computer Sciences.

\newpage

\appendix

\begin{widetext}

\section*{Appendix}

Let $H_0$ be the Hamiltonian capturing the three-band $\mathbf{k}\cdot\mathbf{p}$ electronic band structure in the growth direction ($z$) of the QCL and the free-particle motion in the $x$-$y$ plane. Let $H_i$ represent nonradiative electron interactions (with longitudinal acoustic and optical phonons, interface roughness, ionized impurities, and random alloy scattering). Finally, let $H'(t)$ stand for the interaction between an electron and the optical field, given by
\begin{equation}
\label{ApEq2}
    H'(t) = \mathrm{q}\pazocal{E}_{ac}z\cos(\omega t),
\end{equation}
where $\mathrm{q}$ is electron charge, $\pazocal{E}_{ac}$ is the field amplitude, $z$ is position in the growth direction, and $\omega$ is the frequency of radiation.

After the phonon degrees of freedom have been traced out and the Born and Markov approximations have been employed \cite{Breuer2002,Knezevic2013_JCEL_Review}, we obtain the equation of motion (EoM) for the density matrix \cite{Oli_Dissertation,jonasson2017density,jonasson2017density,Jonasson2016_JCEL,Jonasson2016_Photonics}.
\begin{equation}
\begin{split}
\label{ApEq9}
i\hbar\dot{\rho} = \left[H_0 ,\rho\right] + \left[H'(t) ,\rho\right] -\frac{i}{\hbar}\int_0^\infty ds\left[H_i,  \left[e^{-\frac{i}{\hbar}H_0s}H_ie^{+\frac{i}{\hbar}H_0s},\rho(t)\right]\right].
\end{split}
\end{equation}

\noindent The last term will be denoted as $i\hbar \pazocal{D}\rho$, where $\pazocal{D}\rho$ is the dissipator. The derivation of the dissipator for each specific scattering mechanism is explained in detail in \cite{Oli_Dissertation,jonasson2017density}. Finally, the rotating-wave approximation is employed~\cite{Oli_Dissertation,jonasson2017density} and the final equation of motion is Markovian, of the Lindblad form that guarantees positivity (as shown explicitly in Appendix D of \cite{jonasson2017density}). This EoM with the dissipator is written compactly as

\begin{equation}
\begin{split}
\label{ApEq10}
i\hbar\dot{\rho} = \left[H_0 ,\rho\right] + \left[H'(t) ,\rho\right] +i\hbar\pazocal{D}\rho.
\end{split}
\end{equation}

\noindent Equation (\ref{ApEq10}) can be written in the basis $\ket{i,k} = \ket{i} \otimes \ket{k}$, where $\ket{i}$ labels the eigenstates of the 1D Hamiltonian in the growth direction and the continuous parameter $\ket{k}$ is the amplitude of the wave vector in the $x$-$y$ plane (in-plane motion). In this basis,

\begin{equation}
\begin{split}
\label{ApEq11}
i\hbar\dot{\rho}^{E_{\mathbf{k}}}_{i,j}  = \Delta E_{i,j} \rho^{E_{\mathbf{k}}}_{i,j} + \sum_{l,p}\left(H'_{i,l}\rho^{E_\mathbf{k}}_{l,j} - H'_{p,j}\rho^{E_\mathbf{k}}_{i,p}  \right) + i\hbar\left[\pazocal{D}\rho\right]^{E_{\mathbf{k}}}_{i,j}.
\end{split}
\end{equation}
Here, the matrix elements $\braket{i|\hat{A}|j}$ of operator $\hat{A}$ is denoted by $A_{i,j}$. $\Delta E_{i,j} = E_i - E_j$ is the energy spacing between states $i$ and $j$.
Using Eq.~(\ref{ApEq2}), the EoM can be expanded as
\begin{equation}
\begin{split}\label{ApEq12}
i\hbar\dot{\rho}^{E_\mathbf{k}}_{i,j} =  \Delta E_{i,j}\rho^{E_\mathbf{k}}_{i,j} + \mathrm{q}\pazocal{E}_{ac}\cos(\omega t)\sum_{l,p}\left( z_{i,l}\rho^{E_{\mathbf{k}}}_{l,j} - z_{p,j}\rho^{E_{\mathbf{k}}}_{i,p}\right)  + i\hbar\left[\pazocal{D}\rho\right]^{E_\mathbf{k}}_{i,j}.
\end{split}
\end{equation}
The dissipator can be written as a sum of scattering terms proportional to $\rho_{i,j}^{E_\mathbf{k}}$ ($-\gamma_{i,j}^{E_\mathbf{k}}\rho_{i,j}^{E_\mathbf{k}}$) and the reduced dissipator $\left[\bar{\pazocal{D}}\rho\right]_{i,j}^{E_\mathbf{k}}$ where all the matrix elements of $\rho$ other than $\rho_{i,j}^{E_\mathbf{k}}$ play a role ($\left[\pazocal{D}\rho\right]_{i,j}^{E_\mathbf{k}} = -\gamma_{i,j}^{E_\mathbf{k}}\rho_{i,j}^{E_\mathbf{k}} + \left[\bar{\pazocal{D}}\rho\right]_{i,j}^{E_\mathbf{k}}$); with this sign convention, $\gamma_{i,j}$ are positive and have the meaning of simple scattering rates. Using this, Eq.~(\ref{ApEq12}) is written as
\begin{equation}
\begin{split}\label{ApEq13}
i\hbar\dot{\rho}^{E_\mathbf{k}}_{i,j} = \left(\Delta E_{i,j}-i\hbar\gamma_{i,j}^{E_\mathbf{k}}\right)\rho^{E_\mathbf{k}}_{i,j} + \mathrm{q}\pazocal{E}_{ac}\cos(\omega t)\sum_{l,p}\left( z_{i,l}\rho^{E_{\mathbf{k}}}_{l,j} - z_{p,j}\rho^{E_{\mathbf{k}}}_{i,p}\right)  + i\hbar\left[\bar{\pazocal{D}}\rho\right]_{i,j}^{E_\mathbf{k}}.
\end{split}
\end{equation}

\noindent The second term on the right can be broken into two parts as follows
 \begin{equation}
\begin{split}\label{ApEq14}
\sum_{l,p}\left(z_{i,l}\rho^{E_\mathbf{k}}_{l,j} - z_{p,j}\rho^{E_\mathbf{k}}_{i,p}  \right)= \sum_{\substack{l \neq i \\ p \neq j}}\left(z_{i,l}\rho^{E_\mathbf{k}}_{l,j} - z_{p,j}\rho^{E_\mathbf{k}}_{i,p}  \right) + (z_{i,i} - z_{j,j})\rho^{E_\mathbf{k}}_{i,j}.
\end{split}
\end{equation}

\noindent This separation can be employed to write Eq.~(\ref{ApEq13}) as
\begin{equation}
\begin{split}\label{ApEq15}
i\hbar\dot{\rho}^{E_\mathbf{k}}_{i,j} & = \left(\Delta E_{i,j} -i\hbar\gamma_{i,j}^{E_\mathbf{k}}
+ \mathrm{q}\pazocal{E}_{ac}\left(z_{i,i} - z_{j,j}\right)\cos(\omega t)\right)\rho^{E_\mathbf{k}}_{i,j} \\ &
+ \mathrm{q}\pazocal{E}_{ac}\cos(\omega t)\sum_{\substack{l \neq i \\ p \neq j}}\left(z_{i,l}\rho^{E_\mathbf{k}}_{l,j} - z_{p,j}\rho^{E_\mathbf{k}}_{i,p}  \right) + i\hbar\left[\bar{\pazocal{D}}{\rho}\right]^{E_\mathbf{k}}_{i,j}.
\end{split}
\end{equation}

\noindent Equation (\ref{ApEq15}) is solved iteratively. The last two terms on the right-hand side will use the previous iteration of $\rho$, denoted by $\breve\rho$ here, whereby the equation is turned into a first-order ordinary differential equation with a general solution

\begin{equation}
\begin{split}\label{ApEq17}
\rho^{E_\mathbf{k}}_{i,j} = & e^{-i\int \frac{dt}{\hbar} \left(\Delta E_{i,j} -i\hbar\gamma_{i,j}^{E_\mathbf{k}} + V_{i,j}\cos(\omega t)\right)}\times\Biggl\{ \int dt \: \left[\bar{\pazocal{D}}\breve{\rho}\right]^{E_\mathbf{k}}_{i,j}\: e^{i\int \frac{dt}{\hbar} \left(\Delta E_{i,j} -i\hbar\gamma_{i,j}^{E_\mathbf{k}} + V_{i,j}\cos(\omega t)\right)}\\ - &
\frac{i}{\hbar}\mathrm{q}\pazocal{E}_{ac}\sum_{\substack{l \neq i \\ p \neq j}}\int  dt \: \left(z_{i,l}\Breve{\rho}^{E_\mathbf{k}}_{l,j} - z_{p,j}\Breve{\rho}^{E_\mathbf{k}}_{i,p}  \right)\cos(\omega t) \: e^{i\int \frac{dt}{\hbar} \left(\Delta E_{i,j} -i\hbar\gamma_{i,j}^{E_\mathbf{k}} + V_{i,j}\cos(\omega t)\right)}\Biggr\}.
\end{split}
\end{equation}

\noindent Here, $V_{i,j} = \mathrm{q}\pazocal{E}_{ac}\left(z_{i,i} - z_{j,j}\right)$. The first integral on the right-hand side of Eq.~(\ref{ApEq17}) can be simplified as
\begin{equation}
\begin{split}\label{ApEq18}
e^{-i\int \frac{dt}{\hbar} \left(\Delta E_{i,j} -i\hbar\gamma_{i,j}^{E_\mathbf{k}} + V_{i,j}\cos(\omega t)\right)} & = e^{-\frac{i}{\hbar} \left(\Delta E_{i,j}t -i\hbar\gamma_{i,j}^{E_\mathbf{k}}t + V_{i,j}/\omega \sin(\omega t)\right)}\\
& = \sum_n J_n\left(\frac{V_{i,j}}{\hbar \omega}\right)e^{-\frac{i}{\hbar} \left(\Delta E_{i,j} -i\hbar\gamma_{i,j}^{E_\mathbf{k}} + \hbar\omega n \right)t},
\end{split}
\end{equation}
where we have used $e^{ixsin(\theta)} = \sum_n J_n(x)e^{in\theta}$.

The dissipator is linear, so its action can be represented in terms of its actions on the Fourier components of $\breve\rho$:

\begin{equation}
\left[\bar{\pazocal{D}}\breve\rho(t)\right]^{E_\mathbf{k}}_{i,j} = \left[\bar{\pazocal{D}}\sum_q\breve\rho_q e^{iq\omega t}\right]^{E_\mathbf{k}}_{i,j} = \sum_q\left[\bar{\pazocal{D}}\breve\rho_q \right]^{E_\mathbf{k}}_{i,j}e^{iq\omega t}.
\end{equation}
This helps us simplify the second term in the first line of Eq.~(\ref{ApEq17}) in a manner similar to Eq. (\ref{ApEq18}):
\begin{equation}
\begin{split}\label{ApEq20}
\int dt & \: \left[\bar{\pazocal{D}}\breve\rho(t)\right]^{E_\mathbf{k}}_{i,j}\: e^{i\int \frac{dt}{\hbar} \left(\Delta E_{i,j} -i\hbar\gamma_{i,j}^{E_\mathbf{k}} + V_{i,j}\cos(\omega t)\right)} = \\ = & -i\hbar\sum_{m,q}J_{m}\left(\frac{V_{i,j}}{h\omega}\right)\left[\bar{\pazocal{D}}\breve\rho_q\right]^{E_\mathbf{k}}_{i,j}\times \frac{e^{\frac{i}{\hbar} \left(\Delta E_{i,j} -i\hbar\gamma_{i,j}^{E_\mathbf{k}} + \hbar\omega (m+q) \right)t}}{\Delta E_{i,j} -i\hbar\gamma_{i,j}^{E_\mathbf{k}} + \hbar\omega (m+q)}.
\end{split}\end{equation}

\noindent Using Eqs. (\ref{ApEq20}) and (\ref{ApEq18}), one can write the first line of Eq.~(\ref{ApEq17}) as
\begin{equation}
\label{ApEq21}
-i\hbar\sum_{n,m,q}J_n\left(\frac{V_{i,j}}{h\omega}\right)J_{m}\left(\frac{V_{i,j}}{h\omega}\right)\left[\bar{\pazocal{D}}\breve\rho_q\right]^{E_\mathbf{k}}_{i,j}\times \frac{e^{-i(n-m-q)\omega t}}{\Delta E_{i,j} -i\hbar\gamma_{i,j}^{E_\mathbf{k}} + \hbar\omega (m+q)}.
\end{equation}


\noindent The second line in Eq. (\ref{ApEq17}) can be rewritten as:

\begin{equation}
\begin{split}
 &-\frac{i}{\hbar}\mathrm{q}\pazocal{E}_{ac}\sum_{\substack{l \neq i \\ p \neq j}}e^{-i\int \frac{dt}{\hbar} \left(\Delta E_{i,j} -i\hbar\gamma_{i,j}^{E_\mathbf{k}} + V_{i,j}cos(\omega t)\right)} \\
&\times \int  dt \left(z_{i,l}{\breve\rho}^{E_\mathbf{k}}_{l,j} - z_{p,j}{\breve\rho}^{E_\mathbf{k}}_{i,p}  \right)\: \cos(\omega t) \: e^{i\int \frac{dt}{\hbar} \left(\Delta E_{i,j} -i\hbar\gamma_{i,j}^{E_\mathbf{k}} + V_{i,j}cos(\omega t)\right)}.
\end{split}
\end{equation}

\noindent Using $\cos(\omega t) = \frac{1}{2}(e^{i\omega t} + e^{-i\omega t})$ and writing the density matrices as a summation of their harmonics:

\begin{equation}
\begin{split}
 & -\frac{i}{2\hbar}\mathrm{q}\pazocal{E}_{ac}\sum_{n,q}\sum_{\substack{l \neq i \\ p \neq j}}J_n\left(\frac{V_{i,j}}{\hbar \omega}\right)e^{-i\int \frac{dt}{\hbar} \left(\Delta E_{i,j} -i\hbar\gamma_{i,j}^{E_\mathbf{k}} + V_{i,j}cos(\omega t)\right)} \\
 &\times \int  dt \left(z_{i,l}{[\breve\rho_q]}^{E_\mathbf{k}}_{l,j} - z_{p,j}{[\breve\rho_q]}^{E_\mathbf{k}}_{i,p}  \right)\:\left(e^{i\omega(1+q) t} + e^{-i\omega(1-q) t}\right) \: e^{+\frac{i}{\hbar} \left(\Delta E_{i,j} -i\hbar\gamma_{i,j}^{E_\mathbf{k}} + \hbar\omega n \right)t}.
\end{split}
\end{equation}

\noindent Combining the exponential terms,
\begin{equation}
\begin{split}
 &-\frac{i}{2\hbar}\mathrm{q}\pazocal{E}_{ac}\sum_{n,q}\sum_{\substack{l \neq i \\ p \neq j}}\left(z_{i,l}{[\breve\rho_q]}^{E_\mathbf{k}}_{l,j} - z_{p,j}{[\breve\rho_q]}^{E_\mathbf{k}}_{i,p}  \right)\: J_n\left(\frac{V_{i,j}}{\hbar \omega}\right)e^{-i\int \frac{dt}{\hbar} \left(\Delta E_{i,j} -i\hbar\gamma_{i,j}^{E_\mathbf{k}} + V_{i,j}cos(\omega t)\right)} \\
 &\times \int  dt \:
\left( e^{+\frac{i}{\hbar} \left(\Delta E_{i,j} -i\hbar\gamma_{i,j}^{E_\mathbf{k}} + \hbar\omega (n+q+1) \right)t} + e^{+\frac{i}{\hbar} \left(\Delta E_{i,j} -i\hbar\gamma_{i,j}^{E_\mathbf{k}} + \hbar\omega (n+q-1) \right)t}  \right).
\end{split}
\end{equation}

\noindent After taking the integrals,

\begin{equation}
\begin{split}
 &-\frac{\mathrm{q}\pazocal{E}_{ac}}{2}\sum_{n,m,q}\sum_{\substack{l \neq i \\ p \neq j}}\left(z_{i,l}{[\breve\rho_q]}^{E_\mathbf{k}}_{l,j} - z_{p,j}{[\breve\rho_q]}^{E_\mathbf{k}}_{i,p}  \right)\: J_n\left(\frac{V_{i,j}}{\hbar \omega}\right)J_m\left(\frac{V_{i,j}}{\hbar \omega}\right)e^{-\frac{i}{\hbar} \left(\Delta E_{i,j} -i\hbar\gamma_{i,j}^{E_\mathbf{k}} + \hbar\omega m \right)t} \\
& \times
\left( \frac{e^{+\frac{i}{\hbar} \left(\Delta E_{i,j} -i\hbar\gamma_{i,j}^{E_\mathbf{k}} + \hbar\omega (n+q+1) \right)t}}{\Delta E_{i,j} -i\hbar\gamma_{i,j}^{E_\mathbf{k}} + \hbar\omega (n+q+1)} + \frac{e^{+\frac{i}{\hbar} \left(\Delta E_{i,j} -i\hbar\gamma_{i,j}^{E_\mathbf{k}} + \hbar\omega (n+q-1) \right)t}}{\Delta E_{i,j} -i\hbar\gamma_{i,j}^{E_\mathbf{k}} + \hbar\omega (n+q-1)} \right).
\end{split}
\end{equation}

\noindent Upon combining the exponential terms,

\begin{equation}
\begin{split}
 & -\frac{\mathrm{q}\pazocal{E}_{ac}}{2}\sum_{n,m,q}\sum_{\substack{l \neq i \\ p \neq j}}\left(z_{i,l}{[\breve\rho_q]}^{E_\mathbf{k}}_{l,j} - z_{p,j}{[\breve\rho_q]}^{E_\mathbf{k}}_{i,p}  \right)\: J_n\left(\frac{V_{i,j}}{\hbar \omega}\right)J_m\left(\frac{V_{i,j}}{\hbar \omega}\right) \\
 &\times
\left( \frac{e^{+\frac{i}{\hbar} \hbar\omega (n-m+q+1) t}}{\Delta E_{i,j} -i\hbar\gamma_{i,j}^{E_\mathbf{k}} + \hbar\omega (n+q+1)} + \frac{e^{+\frac{i}{\hbar} \hbar\omega (n-m+q-1) t}}{\Delta E_{i,j} -i\hbar\gamma_{i,j}^{E_\mathbf{k}} + \hbar\omega (n+q-1)} \right).
\end{split}
\end{equation}

\noindent For brevity, it is useful to define a polarization matrix element $\pazocal{A}_{i,j}^{q,E_\mathbf{k}}$ as:

\begin{equation}\label{ApEqPolarization}
    \left[\pazocal{A}_q\right]_{i,j}^{E_\mathbf{k}} =  -\frac{\mathrm{q}\pazocal{E}_{ac}}{2}\sum_{\substack{l \neq i \\ p \neq j}}\left(z_{i,l}{[\breve\rho_q]}^{E_\mathbf{k}}_{l,j} - z_{p,j}{[\breve\rho_q]}^{E_\mathbf{k}}_{i,p}  \right).
\end{equation}


\noindent Overall, Eq.~(\ref{ApEq17}) can be written as:

\begin{equation}
\begin{split}\label{ApEq23}
& \rho_{i,j}^{E_\mathbf{k}}(t) = \sum_{n,m,q}J_n\left(\frac{V_{i,j}}{h\omega}\right)J_{m}\left(\frac{V_{i,j}}{h\omega}\right)\Biggl\{ -i\hbar\left[\bar{\pazocal{D}}\breve\rho_q\right]^{E_\mathbf{k}}_{i,j}\frac{e^{i (n-m+q)\omega t}}{\Delta E_{i,j} -i\hbar\gamma_{i,j}^{E_\mathbf{k}} + \hbar\omega (n+q)}\\
   &+ \left[\pazocal{A}_q\right]_{i,j}^{E_\mathbf{k}}
\left( \frac{e^{i (n-m+q+1) \omega t}}{\Delta E_{i,j} -i\hbar\gamma_{i,j}^{E_\mathbf{k}} + \hbar\omega (n+q+1)} + \frac{e^{ i(n-m+q-1) \omega t}}{\Delta E_{i,j} -i\hbar\gamma_{i,j}^{E_\mathbf{k}} + \hbar\omega (n+q-1)} \right)\Biggr\}.
\end{split}
\end{equation}

 \noindent Equation (\ref{ApEq23}) is the form we solve iteratively. The left-hand side is the updated $\rho$ while everything on the right-hand side depends on the previous iteration $\breve\rho$. To start the iteration, a zeroth-order guess is needed. A simple, intuitive choice is a diagonal density matrix with level occupations corresponding to the Fermi distribution, which we successfully employed here. Alternatively, the steady-state solution without the optical field could be used as a zeroth-order guess; this guess is more involved to compute, but it speeds up the iterative process. Note that the solution (\ref{ApEq23}) is nonperturbative in terms of $\pazocal{E}_{ac}$ and is therefore applicable to high optical-field amplitudes.

\end{widetext}

%

\end{document}